\documentclass[sigchi-a]{acmart}
\usepackage[utf8x]{inputenc}

\def\BibTeX{{\rm B\kern-.05em{\sc i\kern-.025em b}\kern-.08emT\kern-.1667em\lower.7ex\hbox{E}\kern-.125emX}}
    
\copyrightyear{2019}
\acmYear{2019}
\setcopyright{none}
\acmConference[CHI 2019 HCML Workshop]{CHI '19: ACM CHI Conference on Human Factors in Computing Systems}{May 04--09, 2019}{Glasgow, UK}
\acmBooktitle{CHI 2019 HCML Workshop: ACM CHI Conference on Human Factors in Computing Systems}
\acmPrice{15.00}

\begin{document}

%
\title{Designing for the Long Tail of Machine Learning}

%
\author{Martin Lindvall}
\email{martin.lindvall@liu.se}
\affiliation{%
  \institution{Sectra AB}
  \streetaddress{Teknikringen 20}
  \city{Linköping}
  \country{Sweden}
  \postcode{59040}
}
\affiliation{%
  \institution{Department of Science and Technology, ITN, Linköping University}
}

\author{Jesper Molin}
\email{jesper.molin@sectra.com}
\affiliation{%
  \institution{Sectra AB}
  \streetaddress{Teknikringen 20}
  \city{Linkoping}
  \country{Sweden}
  \postcode{59040}
}

%

\begin{abstract}
Recent technical advances has made machine learning (ML) a promising component to include in end user facing systems. However, user experience (UX) practitioners face challenges in relating ML to existing user-centered design processes and how to navigate the possibilities and constraints of this design space. Drawing on our own experience, we characterize designing within this space as navigating trade-offs between data gathering, model development and designing valuable interactions for a given model performance. We suggest that the theoretical description of how machine learning performance scales with training data can guide designers in these trade-offs as well as having implications for prototyping. We exemplify the learning curve's usage by arguing that a useful pattern is to design an initial system in a bootstrap phase that aims to exploit the training effect of data collected at increasing orders of magnitude.
%
\end{abstract}

%

%

%
\maketitle

\section{Introduction}
Recent works have advocated making machine learning (ML) more accessible by helping non-ML experts build and design better learning-based systems \cite{yang_grounding_2018}. The designers of ML-based systems has the potential to improve the experiential value at all stages of development, from problem framing to maintenance phases \cite{gillies_human-centred_2016} but this technology does not come without challenges. Framing ML as a design material highlights the fact that designers must be aware of its properties when used in human-centered design methods \cite{dove_ux_2017}. 

\begin{marginfigure}
  \centering
  \includegraphics[width=\linewidth]{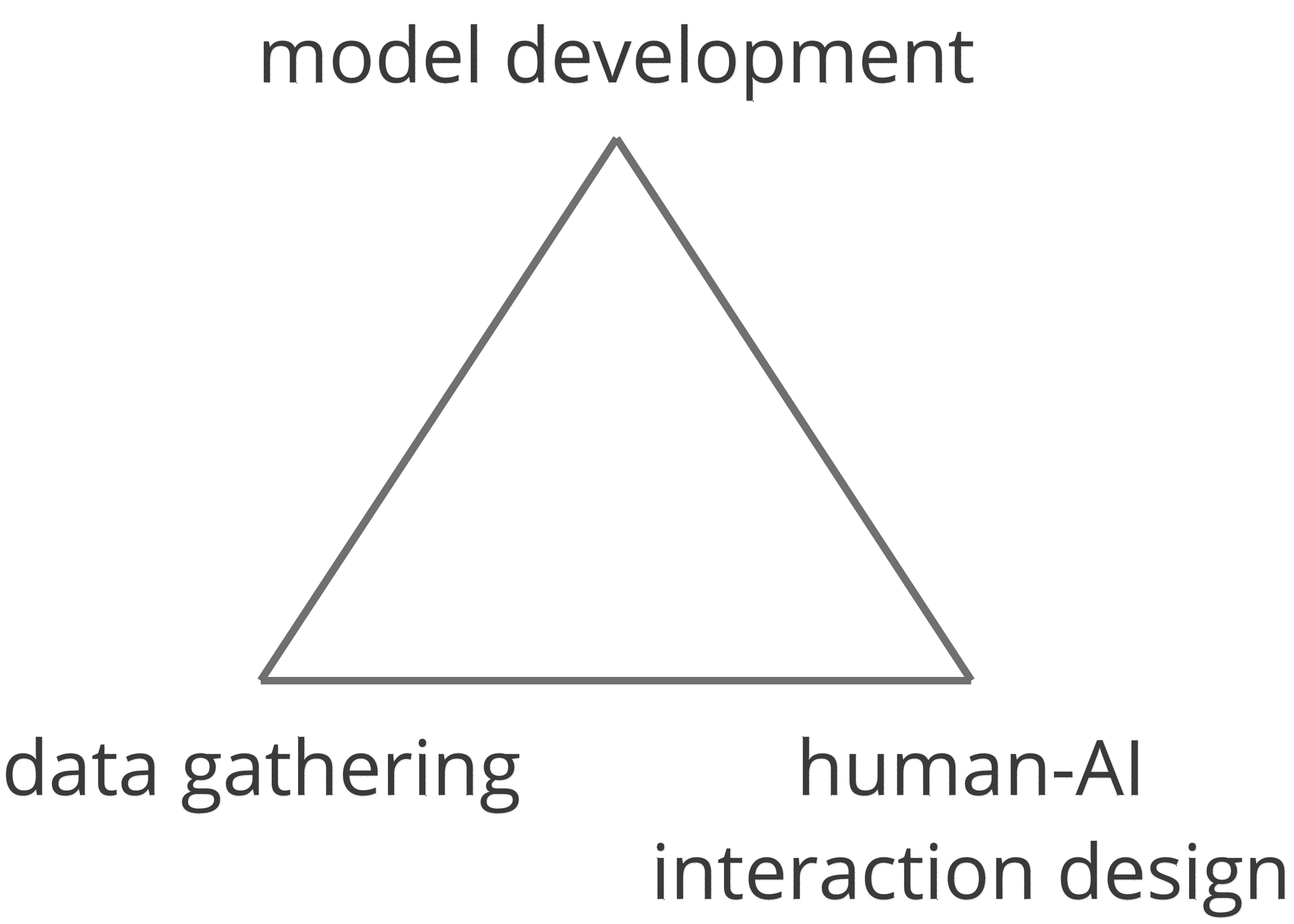}
  \caption{When exploring the design space of a ML-based system, there is usually a trade off between pursuing training algorithm improvements through model development, gathering more training data and attempting to design a suitable interaction using the current best performance.}
  \Description{description.}
  \label{fig:activity_tradeoff}
\end{marginfigure}

In our own practice of prototyping, building and deploying ML-based systems we have often found ourselves trying to decide which activity might at any given time best forward our ambition to create ML-based systems that improve patient outcomes in real clinical situations. As it happens, we are often trying to decide whether to gather more or different training data, to spend time on improving the training algorithms or whether to employ human-centered methods to design user interactions that could render a model's performance usable in a collaborative way within the deployed solution, a trade off we depict schematically in Figure \ref{fig:activity_tradeoff}. While the reality of practice is messy, we hope that by this simple description of three highly ML-related activities we can begin a discourse on how ML-specific constraints and possibilities affect the design process.

\subsection{Navigating the design space of ML}
For our projects to date we have largely relied on what has sometimes been referred to as traditional machine learning \cite{amershi_power_2014}. The part of the process relevant to this discussion of ML typically begins in the corner of what we, for lack of a better term, call \emph{human-AI interaction design}. At some early point, one or multiple rather vague concepts are ideated with end users. Such a process usually results in a problem definition, an idea of how ML might bring value and a very rough sketch of how that value might be realized in use through interaction with end users. Some very quick and dirty conceptual \emph{model development} follow in order to clarify what kind of training data to gather before a very small \emph{data gathering} pilot starts. After some added \emph{model development} on that initial data, the order of the process gets murky. How much model development is worth doing on a small amount of initial data? Should one blindly go collect more data, if so, how much? Furthermore, at which point should we revisit our \emph{human-AI interaction design} to align concepts with new notions of achievable model performances?

Instead of prescribing some order of activities, we think it sensible that for each iteration one is able to reflect upon the cost of each activity in relation to its predicted impact on advancing the overall design goal. While designers probably have a good idea of the costs and benefits of employing user-centered design methods to further \emph{human-AI interaction designs} given a fixed model performance, they might lack tools for estimating the effect of \emph{data gathering} and \emph{model development} on model performance.

As we have continued to work in design of ML-based systems, the theoretical relationship between training data amount and model performance has helped guide our decisions on which part of the improvement triangle to address at which phase of the design project. Recent work has highlighted that UX designers may not have a clear understanding of the relationship ML has with data \cite{dove_ux_2017}, thus we think that an extended discussion on how this relationship impacts design decisions might be generative in informing both process and particular system designs.

\section{The machine learning curve for designers}
Machine learning algorithms train a prediction model from samples. In general, the performance of the trained model improves with the amount and quality of training data. Since the model learns by incorporating new information, the value of each training sample will decline because, if drawn from the same source, chances increase that the new sample embodies something the model has already learned. This means that the generic shape of a learning curve follows an inverse power law, with the details depending on e.g. model size, problem difficulty, and label noise \cite{hestness2017deep}. It has been shown that this power law holds up to at least 300 millions samples, as long as the accuracy level does not reach the inherent noise level in the data \cite{sun2017revisiting}.

\begin{marginfigure}
  \centering
  \includegraphics[width=\linewidth]{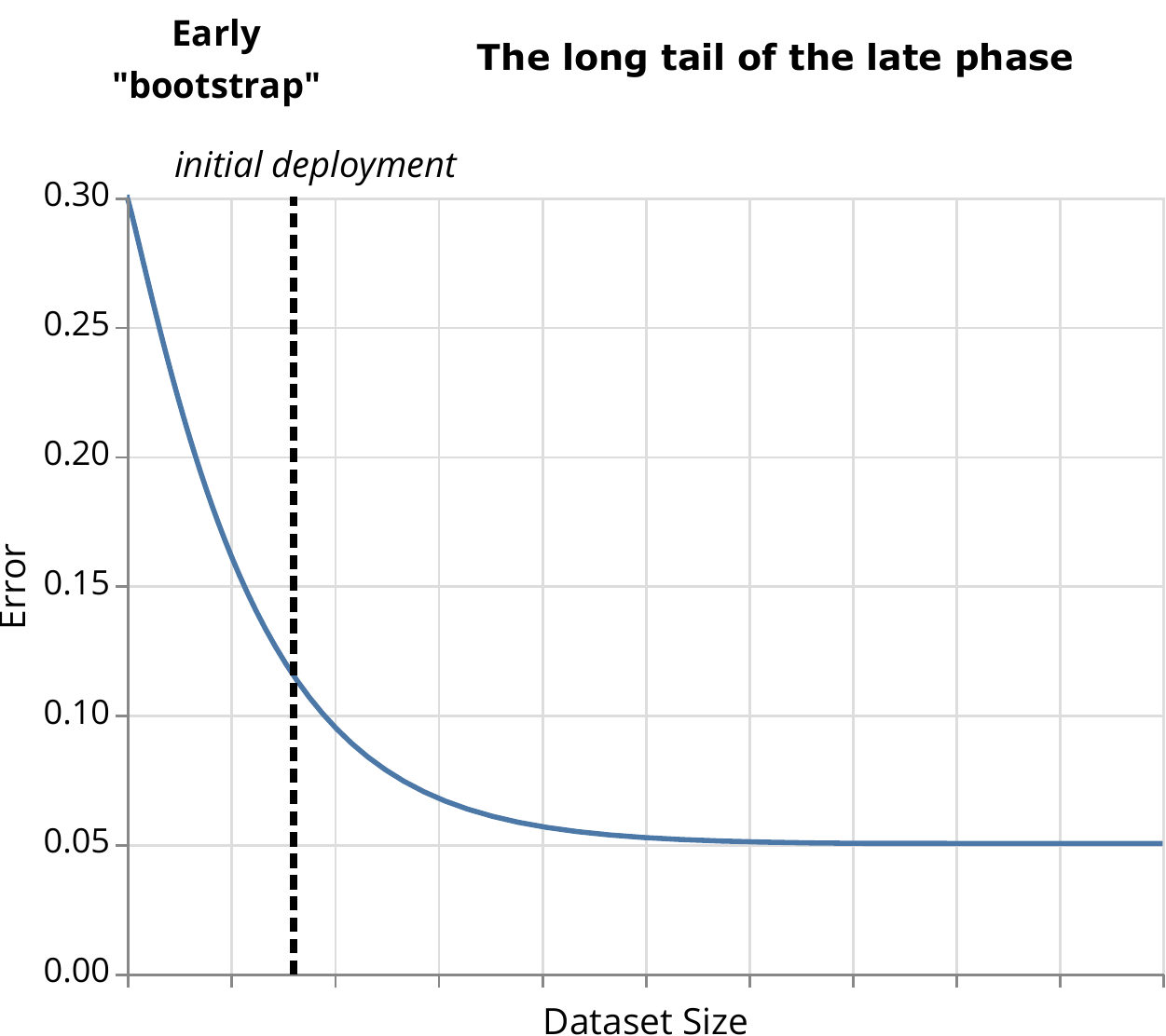}
  \caption{Deep learning model error decreases with increasing data set size following an inverse power law. When designing systems using these models, this can be exploited by dealing with design challenges in the initial \emph{bootstrap} phase and the following \emph{long tail} phase.}
  \Description{}
  \label{fig:longtail}
\end{marginfigure}

As we have previously described \cite{lindvall_machine_2018} we found that a reoccurring theme for our work with creating ML-based systems was first training a model on a relatively small dataset that had been expert-annotated by manual means, to then design an interaction embodying a successful task decomposition such that the collaborative workflow both helps the medical practitioner in her daily work while also implicitly generating training data that we imagined would increase the model's performance by a future update.

This way of approaching system design in two phases can be viewed as exploiting the learning curve to address the cost of gathering data. Since the improvement to the model solely from data requires the collection of data at increasing order of magnitudes, the cost of data collection in the development environment might grow out of proportion quickly. Inversely, trying to deploy a product with too weak a model might not provide enough value to end users to achieve wanted usage, which can be the rationale behind investing in "manual" data gathering solely for development. The relationship between training data amount and model performance and its evolution over time, loosely divided into an early \emph{bootstrap} and a late \emph{long tail} phase, with different characteristics and implications for design is illustrated in Figure \ref{fig:longtail}.

\subsection{Bootstrap phase}
At this stage, the model improves rapidly with relatively small amounts of training data and challenges include problem framing, feature engineering and training a model with sufficient performance to enable realistic user interface prototyping. It is at this stage that the design of manual annotation and labeling tools might be employed to engage domain experts as providers of training labels for the prototype models. The reasonable target performance in the bootstrap phase is probably less than that of the domain experts providing supervised labels, if such are employed. Hence, the \emph{human-AI interaction design} will need to especially consider that the error characteristics of the resulting model predictions can be compensated by human behaviour in the context of use.

\subsection{The long tail}
When starting to approach the long tail, for many practical applications, further improvement to the model performance requires either the data collection to move towards large scale collection or attempting to improve the model by pursuing breakthrough research within machine learning. The design activities will focus on ways to make systems continuously collect training data in a way that it has minimal impact on end users goals and user experience. In order to increase data collection by orders of magnitude, the collection will move from being explicit to implicit. Other notable issues the system design needs to address are model drift, quality assurance, systematic bias and generalizability between contexts.

\section{Impact for human-centered machine learning}
Previous work has indicated that designers believe it may require an "unwieldy amount" of data to create functional prototypes \cite{dove_ux_2017}. The theory surrounding deep learning suggests it is possible to estimate the scaling of model performance with data if enough is gathered to get past a 'small data' plateau and into the power-law region \cite{hestness2017deep}. We believe that by combining experiments using small data and using the learning curve as a heuristic, designers can start imagining the points on the curve where resources required for data collection and model development exceed those that would be required to realize a system enabling valuable interactions with the current model performance. The designer could diverge to a few hypothetical designs by aiming for different target performance levels, as illustrated in figure 3. If the prediction goal of those concepts are kept somewhat similar, this can be done with a low impact in terms of extra resources for  data gathering and model development.

\begin{marginfigure}
  \centering
  \includegraphics[width=\linewidth]{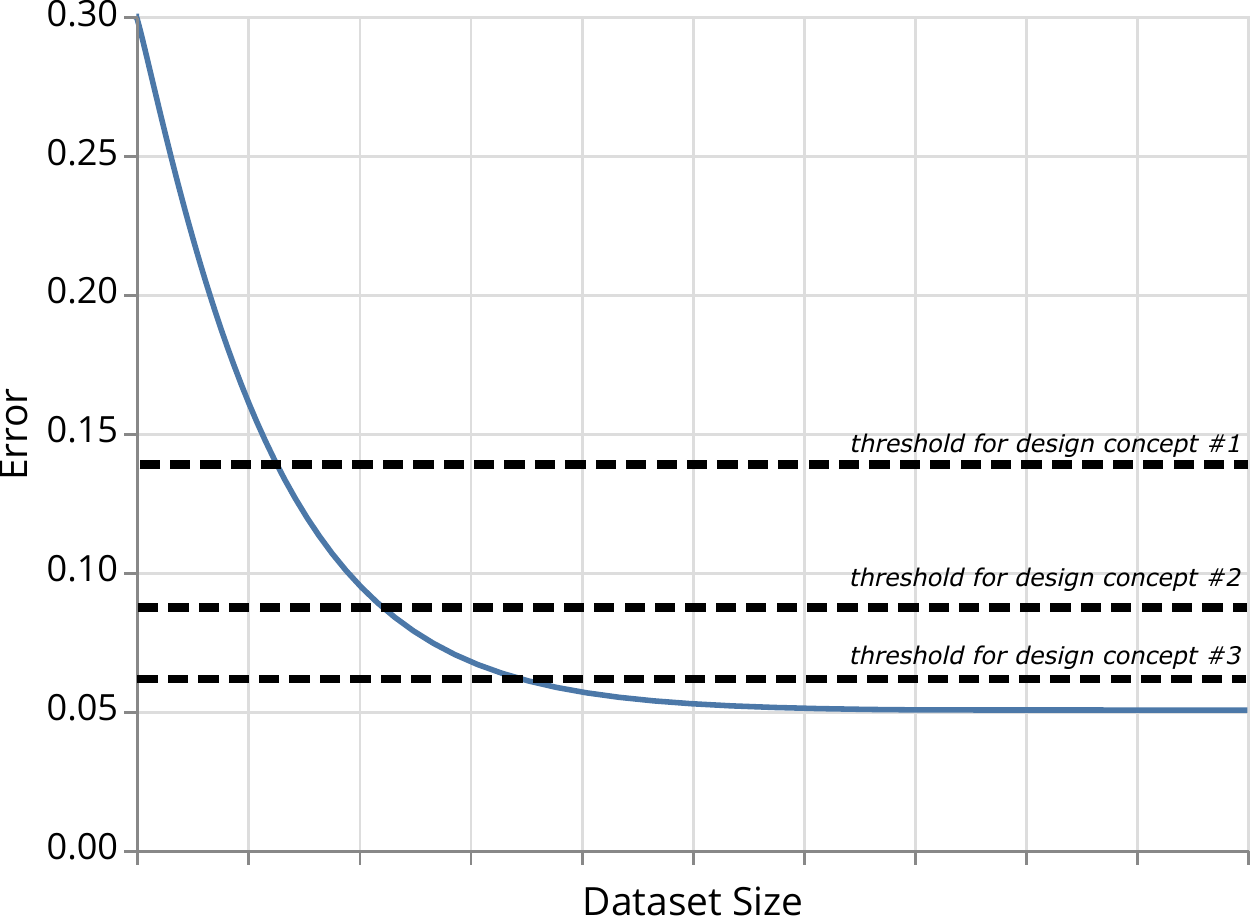}
  \caption{Designers can use the machine learning curve as a tool to diverge multiple hypothetical design concepts that depend on model performance of a certain quality.}
  \Description{}
  \label{fig:destargets}
\end{marginfigure}

By creating a timeline for ML-based systems associated with both machine performance and amount of training data, it becomes possible to nuance questions such as \emph{“are human labelers useful or should we do something different?”}. By the principles we have introduced we might conclude that such model-centered performance activities can be useful in the bootstrap phase but with declining motivations in later phases.

The learning curve relates the effect of additional training data to model performance with a fixed model training pipeline. However, activities of model development such as algorithm selection, hyper parameter tuning and feature engineering can also effect model performance. In our practice, we constrain our explorations of this vast space of options to applying techniques and pipelines that have been shown to be fruitful in scenarios similar to ours or to be widely usable in general. Our rationale behind this is somewhat tentative and based on two assumptions. First, we assume that similarly to how the benefits of increased data declines, so does the impact of model development in relation to time. Second, we believe that the focus on human-AI interaction and the overall solution means that the value of a few percentage's increase in performance, while being very important to machine learning researchers, is less important to designers.

Finally, the learning curve assumes that examples are uniformly drawn from a population. In interactive machine learning systems \cite{amershi_power_2014} that lets users iteratively refine training data by observing model performance, the subsequent examples might still be very informative. For instance \cite{harvey_user-driven_2016} showed that single-user selected sampling can outperform random sampling. However, the learning curve of this kind of user sampling as seen over large data sets is to our best knowledge, yet unknown.



%
\begin{acks}
This work was partially supported by the Wallenberg AI, Autonomous Systems and Software Program (WASP).
\end{acks}

%
\bibliographystyle{ACM-Reference-Format}
\bibliography{long-tail-main}


\begin{thebibliography}{8}


\ifx \showCODEN    \undefined \def \showCODEN     #1{\unskip}     \fi
\ifx \showDOI      \undefined \def \showDOI       #1{#1}\fi
\ifx \showISBNx    \undefined \def \showISBNx     #1{\unskip}     \fi
\ifx \showISBNxiii \undefined \def \showISBNxiii  #1{\unskip}     \fi
\ifx \showISSN     \undefined \def \showISSN      #1{\unskip}     \fi
\ifx \showLCCN     \undefined \def \showLCCN      #1{\unskip}     \fi
\ifx \shownote     \undefined \def \shownote      #1{#1}          \fi
\ifx \showarticletitle \undefined \def \showarticletitle #1{#1}   \fi
\ifx \showURL      \undefined \def \showURL       {\relax}        \fi
\providecommand\bibfield[2]{#2}
\providecommand\bibinfo[2]{#2}
\providecommand\natexlab[1]{#1}
\providecommand\showeprint[2][]{arXiv:#2}

\bibitem[\protect\citeauthoryear{Amershi, Cakmak, Knox, and Kulesza}{Amershi
  et~al\mbox{.}}{2014}]%
        {amershi_power_2014}
\bibfield{author}{\bibinfo{person}{Saleema Amershi}, \bibinfo{person}{Maya
  Cakmak}, \bibinfo{person}{William~Bradley Knox}, {and} \bibinfo{person}{Todd
  Kulesza}.} \bibinfo{year}{2014}\natexlab{}.
\newblock \showarticletitle{Power to the {People}: {The} {Role} of {Humans} in
  {Interactive} {Machine} {Learning}}.
\newblock \bibinfo{journal}{\emph{AI Magazine}} \bibinfo{volume}{35},
  \bibinfo{number}{4} (\bibinfo{year}{2014}), \bibinfo{pages}{105--120}.
\newblock
\urldef\tempurl%
\url{https://doi.org/10.1609/aimag.v35i4.2513}
\showDOI{\tempurl}


\bibitem[\protect\citeauthoryear{Dove, Halskov, Forlizzi, and Zimmerman}{Dove
  et~al\mbox{.}}{2017}]%
        {dove_ux_2017}
\bibfield{author}{\bibinfo{person}{Graham Dove}, \bibinfo{person}{Kim Halskov},
  \bibinfo{person}{Jodi Forlizzi}, {and} \bibinfo{person}{John Zimmerman}.}
  \bibinfo{year}{2017}\natexlab{}.
\newblock \showarticletitle{{UX} {Design} {Innovation}: {Challenges} for
  {Working} with {Machine} {Learning} as a {Design} {Material}}.
\newblock \bibinfo{journal}{\emph{CHI '17 Proceedings of the 2017 annual
  conference on Human factors in computing systems}} (\bibinfo{year}{2017}),
  \bibinfo{pages}{278--288}.
\newblock
\showISSN{9781450346559}
\urldef\tempurl%
\url{https://doi.org/10.1145/3025453.3025739}
\showDOI{\tempurl}


\bibitem[\protect\citeauthoryear{Gillies, Fiebrink, Tanaka, Garcia, Bevilacqua,
  Heloir, Nunnari, Mackay, Amershi, Lee, d'Alessandro, Tilmanne, Kulesza, and
  Caramiaux}{Gillies et~al\mbox{.}}{2016}]%
        {gillies_human-centred_2016}
\bibfield{author}{\bibinfo{person}{Marco Gillies}, \bibinfo{person}{Rebecca
  Fiebrink}, \bibinfo{person}{Atau Tanaka}, \bibinfo{person}{Jérémie Garcia},
  \bibinfo{person}{Frédéric Bevilacqua}, \bibinfo{person}{Alexis Heloir},
  \bibinfo{person}{Fabrizio Nunnari}, \bibinfo{person}{Wendy Mackay},
  \bibinfo{person}{Saleema Amershi}, \bibinfo{person}{Bongshin Lee},
  \bibinfo{person}{Nicolas d'Alessandro}, \bibinfo{person}{Joëlle Tilmanne},
  \bibinfo{person}{Todd Kulesza}, {and} \bibinfo{person}{Baptiste Caramiaux}.}
  \bibinfo{year}{2016}\natexlab{}.
\newblock \showarticletitle{Human-{Centred} {Machine} {Learning}}. In
  \bibinfo{booktitle}{\emph{Proceedings of the 2016 {CHI} {Conference}
  {Extended} {Abstracts} on {Human} {Factors} in {Computing} {Systems}}}
  \emph{(\bibinfo{series}{{CHI} {EA} '16})}. \bibinfo{publisher}{ACM},
  \bibinfo{address}{New York, NY, USA}, \bibinfo{pages}{3558--3565}.
\newblock
\showISBNx{978-1-4503-4082-3}
\urldef\tempurl%
\url{https://doi.org/10.1145/2851581.2856492}
\showDOI{\tempurl}
\newblock
\shownote{event-place: San Jose, California, USA.}


\bibitem[\protect\citeauthoryear{Harvey and Porter}{Harvey and Porter}{2016}]%
        {harvey_user-driven_2016}
\bibfield{author}{\bibinfo{person}{Neal Harvey} {and} \bibinfo{person}{Reid
  Porter}.} \bibinfo{year}{2016}\natexlab{}.
\newblock \showarticletitle{User-driven sampling strategies in image
  exploitation}.
\newblock \bibinfo{journal}{\emph{Information Visualization}}
  \bibinfo{volume}{15}, \bibinfo{number}{1} (\bibinfo{date}{Jan.}
  \bibinfo{year}{2016}), \bibinfo{pages}{64--74}.
\newblock
\showISSN{1473-8716}
\urldef\tempurl%
\url{https://doi.org/10.1177/1473871614557659}
\showDOI{\tempurl}


\bibitem[\protect\citeauthoryear{Hestness, Narang, Ardalani, Diamos, Jun,
  Kianinejad, Patwary, Ali, Yang, and Zhou}{Hestness et~al\mbox{.}}{2017}]%
        {hestness2017deep}
\bibfield{author}{\bibinfo{person}{Joel Hestness}, \bibinfo{person}{Sharan
  Narang}, \bibinfo{person}{Newsha Ardalani}, \bibinfo{person}{Gregory Diamos},
  \bibinfo{person}{Heewoo Jun}, \bibinfo{person}{Hassan Kianinejad},
  \bibinfo{person}{Md Patwary}, \bibinfo{person}{Mostofa Ali},
  \bibinfo{person}{Yang Yang}, {and} \bibinfo{person}{Yanqi Zhou}.}
  \bibinfo{year}{2017}\natexlab{}.
\newblock \showarticletitle{Deep learning scaling is predictable, empirically}.
\newblock \bibinfo{journal}{\emph{arXiv preprint arXiv:1712.00409}}
  (\bibinfo{year}{2017}).
\newblock


\bibitem[\protect\citeauthoryear{Lindvall, Molin, and Löwgren}{Lindvall
  et~al\mbox{.}}{2018}]%
        {lindvall_machine_2018}
\bibfield{author}{\bibinfo{person}{Martin Lindvall}, \bibinfo{person}{Jesper
  Molin}, {and} \bibinfo{person}{Jonas Löwgren}.}
  \bibinfo{year}{2018}\natexlab{}.
\newblock \showarticletitle{From {Machine} {Learning} to {Machine} {Teaching}:
  {The} {Importance} of {UX}}.
\newblock \bibinfo{journal}{\emph{Interactions}} \bibinfo{volume}{25},
  \bibinfo{number}{6} (\bibinfo{date}{Oct.} \bibinfo{year}{2018}),
  \bibinfo{pages}{52--57}.
\newblock
\showISSN{1072-5520}
\urldef\tempurl%
\url{https://doi.org/10.1145/3282860}
\showDOI{\tempurl}


\bibitem[\protect\citeauthoryear{Sun, Shrivastava, Singh, and Gupta}{Sun
  et~al\mbox{.}}{2017}]%
        {sun2017revisiting}
\bibfield{author}{\bibinfo{person}{Chen Sun}, \bibinfo{person}{Abhinav
  Shrivastava}, \bibinfo{person}{Saurabh Singh}, {and} \bibinfo{person}{Abhinav
  Gupta}.} \bibinfo{year}{2017}\natexlab{}.
\newblock \showarticletitle{Revisiting unreasonable effectiveness of data in
  deep learning era}. In \bibinfo{booktitle}{\emph{Computer Vision (ICCV), 2017
  IEEE International Conference on}}. IEEE, \bibinfo{pages}{843--852}.
\newblock


\bibitem[\protect\citeauthoryear{Yang, Suh, Chen, and Ramos}{Yang
  et~al\mbox{.}}{2018}]%
        {yang_grounding_2018}
\bibfield{author}{\bibinfo{person}{Qian Yang}, \bibinfo{person}{Jina Suh},
  \bibinfo{person}{Nan-Chen Chen}, {and} \bibinfo{person}{Gonzalo Ramos}.}
  \bibinfo{year}{2018}\natexlab{}.
\newblock \showarticletitle{Grounding {Interactive} {Machine} {Learning} {Tool}
  {Design} in {How} {Non}-{Experts} {Actually} {Build} {Models}}. In
  \bibinfo{booktitle}{\emph{Proceedings of the 2018 {Designing} {Interactive}
  {Systems} {Conference}}} \emph{(\bibinfo{series}{{DIS} '18})}.
  \bibinfo{publisher}{ACM}, \bibinfo{address}{New York, NY, USA},
  \bibinfo{pages}{573--584}.
\newblock
\showISBNx{978-1-4503-5198-0}
\urldef\tempurl%
\url{https://doi.org/10.1145/3196709.3196729}
\showDOI{\tempurl}


\end{thebibliography}

%

\end{document}